\documentclass[10pt,a4paper,onecolumn]{article}
\usepackage{marginnote}
\usepackage{graphicx}
\usepackage{xcolor}
\usepackage{authblk,etoolbox}
\usepackage{titlesec}
\usepackage{calc}
\usepackage{tikz}
\usepackage{hyperref}
\hypersetup{colorlinks,breaklinks,
            urlcolor=[rgb]{0.0, 0.5, 1.0},
            linkcolor=[rgb]{0.0, 0.5, 1.0}}
\usepackage{caption}
\usepackage{tcolorbox}
\usepackage{amssymb,amsmath}
\usepackage{ifxetex,ifluatex}
\usepackage{seqsplit}
\usepackage{xstring}

\usepackage{float}
\let\origfigure\figure
\let\endorigfigure\endfigure
\renewenvironment{figure}[1][2] {
    \expandafter\origfigure\expandafter[H]
} {
    \endorigfigure
}

\usepackage{fixltx2e} 

\let\textttOrig=\texttt
\def\texttt#1{\expandafter\textttOrig{\seqsplit{#1}}}
\renewcommand{\seqinsert}{\ifmmode
  \allowbreak
  \else\penalty6000\hspace{0pt plus 0.02em}\fi}

\makeatletter
\let\href@Orig=\href
\def\href@Urllike#1#2{\href@Orig{#1}{\begingroup
    \def\Url@String{#2}\Url@FormatString
    \endgroup}}
\def\href@Notdoi#1#2{\def\tempa{#1}\def\tempb{#2}%
  \ifx\tempa\tempb\relax\href@Urllike{#1}{#2}\else
  \href@Orig{#1}{#2}\fi}
\def\href#1#2{%
  \IfBeginWith{#1}{https://doi.org}%
  {\href@Urllike{#1}{#2}}{\href@Notdoi{#1}{#2}}}
\makeatother

\newlength{\cslhangindent}
\newlength{\csllabelwidth}
\usepackage{calc}

\usepackage[top=3.5cm, bottom=3cm, right=1.5cm, left=1.0cm,
            headheight=2.2cm, reversemp, includemp, marginparwidth=4.5cm]{geometry}

\titleformat{\section}
  {\normalfont\sffamily\Large\bfseries}
  {}{0pt}{}
\titleformat{\subsection}
  {\normalfont\sffamily\large\bfseries}
  {}{0pt}{}
\titleformat{\subsubsection}
  {\normalfont\sffamily\bfseries}
  {}{0pt}{}
\titleformat*{\paragraph}
  {\sffamily\normalsize}

\usepackage{fancyhdr}
\makeatletter
\let\ps@plain\ps@fancy
\fancyheadoffset[L]{4.5cm}
\fancyfootoffset[L]{4.5cm}

\definecolor{linky}{rgb}{0.0, 0.5, 1.0}

\newtcolorbox{repobox}
   {colback=red, colframe=red!75!black,
     boxrule=0.5pt, arc=2pt, left=6pt, right=6pt, top=3pt, bottom=3pt}

\newcommand{\ExternalLink}{%
   \tikz[x=1.2ex, y=1.2ex, baseline=-0.05ex]{%
       \begin{scope}[x=1ex, y=1ex]
           \clip (-0.1,-0.1)
               --++ (-0, 1.2)
               --++ (0.6, 0)
               --++ (0, -0.6)
               --++ (0.6, 0)
               --++ (0, -1);
           \path[draw,
               line width = 0.5,
               rounded corners=0.5]
               (0,0) rectangle (1,1);
       \end{scope}
       \path[draw, line width = 0.5] (0.5, 0.5)
           -- (1, 1);
       \path[draw, line width = 0.5] (0.6, 1)
           -- (1, 1) -- (1, 0.6);
       }
   }

\patchcmd{\@maketitle}{center}{flushleft}{}{}
\patchcmd{\@maketitle}{center}{flushleft}{}{}
\patchcmd{\@maketitle}{\LARGE}{\LARGE\sffamily}{}{}
\def\maketitle{{%
  
  \AB@maketitle}}
\makeatletter
\renewcommand\AB@affilsepx{ \protect\Affilfont}
\renewcommand\AB@affilnote[1]{{\bfseries #1}\hspace{3pt}}
\renewcommand{\affil}[2][]%
   {\newaffiltrue\let\AB@blk@and\AB@pand
      \if\relax#1\relax\def\AB@note{\AB@thenote}\else\def\AB@note{#1}%
        \setcounter{Maxaffil}{0}\fi
        \begingroup
        \let\href=\href@Orig
        \let\texttt=\textttOrig
        \let\protect\@unexpandable@protect
        \def\thanks{\protect\thanks}\def\footnote{\protect\footnote}%
        \@temptokena=\expandafter{\AB@authors}%
        {\def\\{\protect\\\protect\Affilfont}\xdef\AB@temp{#2}}%
         \xdef\AB@authors{\the\@temptokena\AB@las\AB@au@str
         \protect\\[\affilsep]\protect\Affilfont\AB@temp}%
         \gdef\AB@las{}\gdef\AB@au@str{}%
        {\def\\{, \ignorespaces}\xdef\AB@temp{#2}}%
        \@temptokena=\expandafter{\AB@affillist}%
        \xdef\AB@affillist{\the\@temptokena \AB@affilsep
          \AB@affilnote{\AB@note}\protect\Affilfont\AB@temp}%
      \endgroup
       \let\AB@affilsep\AB@affilsepx
}
\makeatother

\renewcommand\Affilfont{\sffamily\small\mdseries}
\ifnum 0\ifxetex 1\fi\ifluatex 1\fi=0 
  \usepackage[T1]{fontenc}
  \usepackage[utf8]{inputenc}

\else 
  \ifxetex
    \usepackage{mathspec}
  \else
    \usepackage{fontspec}
  \fi
  \defaultfontfeatures{Ligatures=TeX,Scale=MatchLowercase}

\fi
\IfFileExists{upquote.sty}{\usepackage{upquote}}{}
\IfFileExists{microtype.sty}{%
\usepackage{microtype}
\UseMicrotypeSet[protrusion]{basicmath} 
}{}

\usepackage{hyperref}
\hypersetup{unicode=true,
            pdftitle={Correlation: An Analyzing Tool for Liquids and for Amorphous Solids},
            pdfborder={0 0 0},
            breaklinks=true}
\let\addcontentslineOrig=\addcontentsline
\def\addcontentsline#1#2#3{\bgroup
  \let\texttt=\textttOrig\addcontentslineOrig{#1}{#2}{#3}\egroup}
\let\markbothOrig\markboth
\def\markboth#1#2{\bgroup
  \let\texttt=\textttOrig\markbothOrig{#1}{#2}\egroup}
\let\markrightOrig\markright
\def\markright#1{\bgroup
  \let\texttt=\textttOrig\markrightOrig{#1}\egroup}

\usepackage{graphicx,grffile}
\makeatletter
\def\maxwidth{\ifdim\Gin@nat@width>\linewidth\linewidth\else\Gin@nat@width\fi}
\def\maxheight{\ifdim\Gin@nat@height>\textheight\textheight\else\Gin@nat@height\fi}
\makeatother
\setkeys{Gin}{width=\maxwidth,height=\maxheight,keepaspectratio}
\IfFileExists{parskip.sty}{%
\usepackage{parskip}
}{
\setlength{\parindent}{0pt}
\setlength{\parskip}{6pt plus 2pt minus 1pt}
}
\ifx\paragraph\undefined\else
\let\oldparagraph\paragraph
\renewcommand{\paragraph}[1]{\oldparagraph{#1}\mbox{}}
\fi
\ifx\subparagraph\undefined\else
\let\oldsubparagraph\subparagraph
\renewcommand{\subparagraph}[1]{\oldsubparagraph{#1}\mbox{}}
\fi

\title{Correlation: An Analyzing Tool for Liquids and for Amorphous
Solids}

        \author[1]{Isaías Rodríguez}
          \author[2]{Renela M. Valladares}
          \author[2]{Alexander Valladares}
          \author[1]{David Hinojosa-Romero}
          \author[3]{Ulises Santiago}
          \author[1]{Ariel A. Valladares}
    
      \affil[1]{Instituto de Investigaciones en Materiales, Universidad
Nacional Autónoma de México}
      \affil[2]{Facultad de Ciencias, Universidad Nacional Autónoma de
México}
      \affil[3]{Department of Computational and Systems Biology,
University of Pittsburgh}
  \date{\vspace{-5ex}}

\begin{document}
\maketitle

\marginpar{
  \sffamily\small

  {\bfseries DOI:} \href{https://doi.org/10.5281/zenodo.4313127}{\color{linky}{10.5281/zenodo.4313127}}

  \vspace{2mm}

  {\bfseries Software}
  \begin{itemize}
    \setlength\itemsep{0em}
    \item \href{}{\color{linky}{Review}} \ExternalLink
    \item \href{https://github.com/Isurwars/Correlation}{\color{linky}{Repository}} \ExternalLink
    \item \href{10.5281/zenodo.4313127}{\color{linky}{Archive}} \ExternalLink
  \end{itemize}

  \vspace{2mm}

  {\bfseries Submitted:} 11 December 2020\\
  {\bfseries Published:} 

  \vspace{2mm}
  {\bfseries License}\\
  Authors of papers retain copyright and release the work under a Creative Commons Attribution 4.0 International License (\href{https://creativecommons.org/licenses/by/4.0/}{\color{linky}{CC BY 4.0}}).
}

\section{Summary}

For almost a century, since Bernal's attempts at a molecular theory of
liquid structure (Bernal \cite{bernal_attempt_1937}), correlation functions have been the
bridge to compare theoretical calculations with experimental
measurements in the study of disordered materials.

Pair Distribution Functions (\(g(r)\)), Radial Distribution Functions
(\(J(r)\)), Plane Angle Distributions (\(g(\theta)\)) and Coordination
Numbers (\(n_c\)) have been widely used to characterize amorphous and
liquid materials (Waseda \cite{waseda_structure_1980}; Elliott \cite{elliott_physics_1986}; Valladares et al. \cite{valladares_new_2011})
and, in particular Bulk Metallic Glasses (Miller and Liaw \cite{miller_bulk_2007};
Galván-Colín et al. \cite{galvan-colin_short-range_2015}).

\textbf{Correlation} is an Open-Source software designed to analyze
liquid structures and amorphous solids; the software is user-friendly,
the modular design makes it easy to integrate in High-Throughput
Computing (HTC) to process structures with a large number of
constituents in a standardized fashion. \textbf{Correlation} is ready to
be used in Windows, Linux and Mac. Currently, we support DMol3 (CAR),
CASTEP (CELL), ONETEP (DAT) and VASP (POSCAR) structure files. The code
can handle up to 25,000 atoms, so it can be used to analyze both
classical and first-principles simulations. At the end, the output of
every single correlation function is exported to the corresponding
comma-separated value file (CSV), to further analyze the results.

\section{Statement of Need}

As time goes by, the number of atoms in theoretical calculations has
grown from a few dozens to hundreds of thousands of atoms, and with this
increment the complexity to calculate the correlation functions that
represent the structure of materials has steadily increased. To answer
this need, there have been several tools developed to calculate some of
the most used correlation functions like: pair distribution functions,
radial distribution functions and plane angle distributions. However,
the use of these tools has been limited, either by a prohibiting cost,
or has been restricted to private academic groups, or geopolitical
limitations introduced by the licensing, or by being specially designed
to an specific software for material simulation.

With these limitations in mind, we decided to create a software that
could calculate several correlation functions of materials, as well as
the more interesting properties derived from these functions. While
making the software accessible to as many people as possible.

\section{Mathematical Background}

\subsection{Pair Distribution Functions}

The structure factor is one of the most useful tools in analyzing the
scattering patterns obtained from X-ray, electron and neutron
diffraction experiments of crystalline, disordered and amorphous
materials. The Fourier transform of the scattering intensity given by
the structure factor \(S(Q)\), yields the pair distribution function
(PDF) \(g(r)\) defined by:

\begin{equation}\label{eq:PDF}
G(r) = 4\pi\rho_0(g(r)-1)=\frac{2}{\pi}\int_{0}^{\infty} Q[S(Q)-1]sin(Qr)dQ,
\end{equation} where \(G(r)\) is the reduced pair distribution function
(rPDF) which will be discussed next.

\begin{figure}
\centering
\includegraphics[width=0.75\textwidth,height=\textheight]{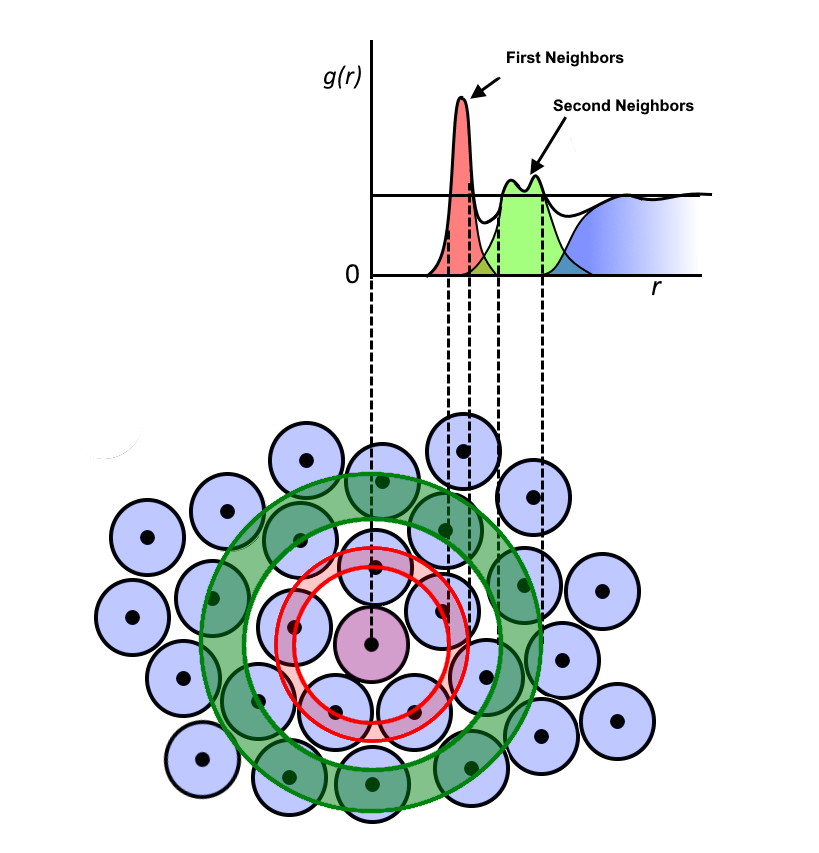}
\caption{Schematic depiction of the first and second neighbor's
coordination spheres for an amorphous metallic alloy and the
corresponding pair distribution function. Design inspired by J. M.
Ziman, Models of disorder (Cambridge University Press \cite{ziman_models_1979}).
\label{fig:RDF}}
\end{figure}

The pair distribution function could also be seen like a distance map
inside the material, the \(g(r)\) function gives how feasible is finding
two atoms separated by the distance (\(r\)) as can be seen in
\ref{fig:RDF} (Ziman 1979).

The PDF is normalized so that, as \(r \to \infty\), \(g(r) \to 1\),
Also, as \(r \to 0\) (for \(r\) shorter than the distance of the closest
approach of pair of atoms), \(g(r)\) becomes zero. The main advantage of
the PDF and the related functions, is that they emphasize the
short-range order of a material.

\subsubsection{\texorpdfstring{Reduced pair distribution function
\(G(r)\)}{Reduced pair distribution function G(r)}}

One of the most widely used pair correlation function is the reduced
pair distribution function. This is defined as
\(G(r) = 4\pi \rho_0 (g(r)-1)\). From this definition, and the
previously discussed tendency at large \(r\) of the PDF, it's clear that
the reduced pair distribution function (rPDF) oscillates around zero as
\(r \to \infty\). It also becomes evident that as \(r \to 0\) (for \(r\)
smaller than the closest pair of atoms), the rPDF behaves like
\(-4\pi \rho_0\).

While the PDF (\(g(r)\)) has an intuitive geometric definition, the rPDF
(\(G(r)\)) can be directly obtained by a Fourier transform of the
structure factor (\(S(Q)\)) as can be seen in \ref{eq:PDF}; this
close relation with the experimental data explains the popularity that
this function has. Also, it has another advantage over other correlation
functions {[}like PDF (\(g(r)\)){]} as the numerical density
\(\rho_0 = N/V\) needs to be estimated in order to normalize the
functions. This is not necessary in rPDF (\(G(r)\)); where this
information is already contained in \(G(r)\) as the slope of the
function when \(r \to 0\).

\subsubsection{\texorpdfstring{Radial distribution function
\(J(r)\)}{Radial distribution function J(r)}}

The last correlation function we shall discuss is also one of the most
physically intuitive, the RDF, \(J(r)\), which is related to the Pair
Distribution Function (PDF) by:

\begin{equation}\label{eq:RDF}
J(r) = 4\pi r^{2} \rho_0 g(r).
\end{equation}

The RDF has the useful property that the quantity \(J(r)dr\) gives the
number of atoms in a spherical shell with inner radius \(r\) and
thickness \(dr\) around every atom as depicted in \ref{fig:RDF}. For
example, the coordination number, or the number of neighbors (\(n_c\)),
is given by:

\begin{equation}\label{eq:Correlation}
n_c = \int_{r_1}^{r_2} J(r) dr,
\end{equation}

where \(r_1\) and \(r_2\) define the RDF peak corresponding to the
coordination shells in question.

\subsection{Plane Angle Distribution}

The use of higher order correlation functions to analyze the structure
of liquids and amorphous solids has been proposed in the literature
(Hafner \cite{hafner_triplet_1982}; Galvan-Colin et al. \cite{galvan-colin_ab_2016}), trying to reproduce the success
obtained by Bernal in the analysis of the structure of liquids (Bernal
\cite{bernal_bakerian_1964}).

In particular, the Plane Angle Distribution, also known as the Bond
Angle Distribution \(f(\theta)\) has been used to characterize the
short-range order of amorphous and liquid structures (Galván-Colín et
al. \cite{galvan-colin_short-range_2015}; Galvan-Colin et al. \cite{galvan-colin_ab_2016}; Mata-Pinzón et al. \cite{mata-pinzon_superconductivity_2016}). Here we
propose to substitute the term ``Bond Angle Distribution'' (BAD), by the
term ``Plane Angle Distribution'' (PAD), also frequently used, since in
condensed matter, proximity does not necessarily imply bonding.

\section{Benchmarks}

\begin{figure}
\centering
\includegraphics[width=0.75\textwidth,height=\textheight]{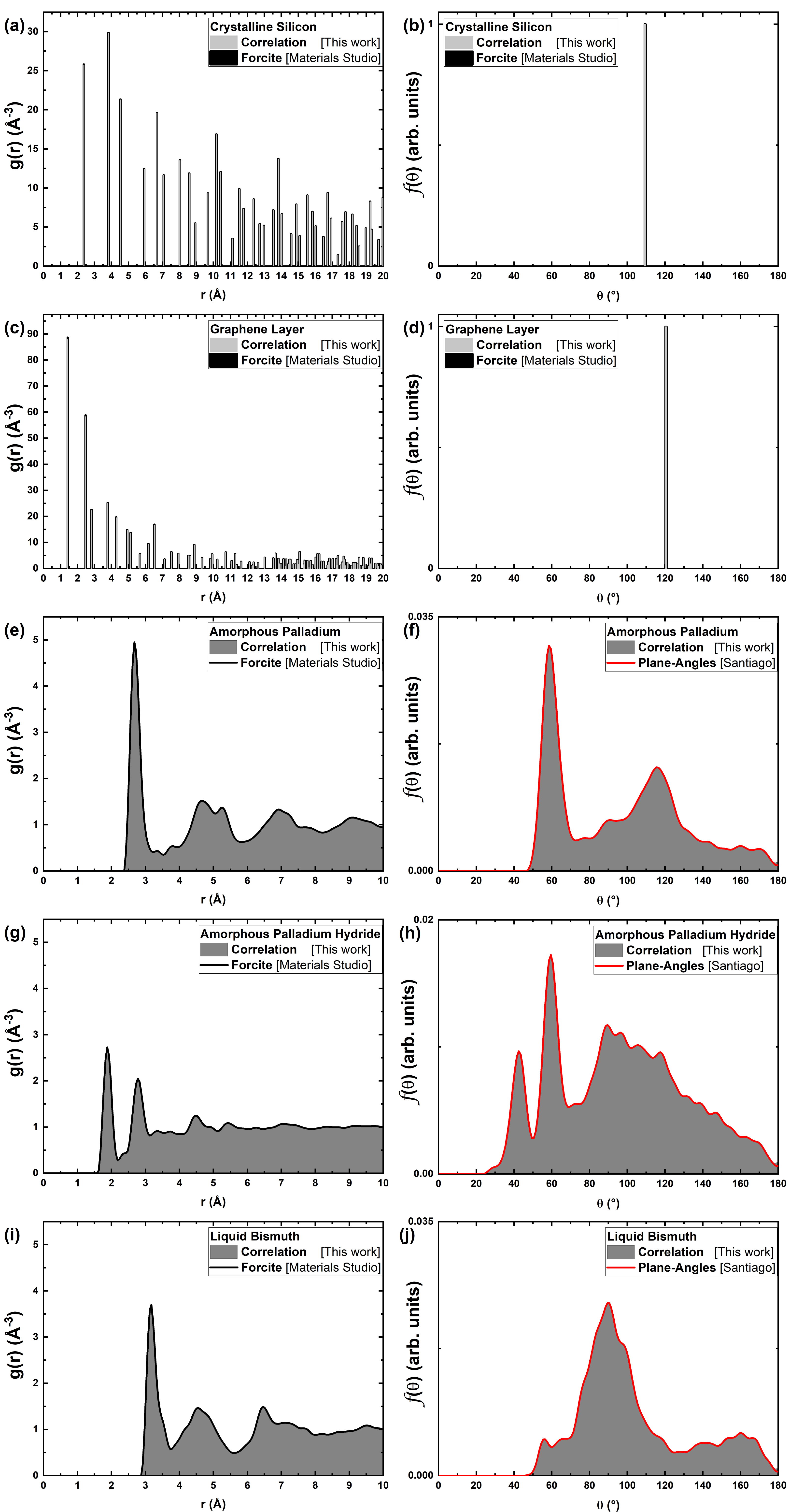}
\caption{Pair Distribution Functions \(g(r)\) on the left, Plane Angle
Distributions on the right for: crystalline silicon, graphene Layer,
amorphous palladium, amorphous palladium hydride and liquid bismuth.
Correlation in gray, Forcite in black, Plane Angles in red. Similarity
is remarkable between \textbf{Correlation} and Forcite as can be seen in
all PDFs. Figures (a) to (d) indicate that the coincidence in the two
results overlap completely. \label{fig:RDF-PAD}}
\end{figure}

In order to assess the performance of \textbf{Correlation}, we
calculated the PDF and PAD for two well known structures (Crystalline
Silicon and a Graphene Layer), and compared the results with the
commercially available software Forcite included in the Materials Studio
suite (``Materials Studio 2016: Forcite, CASTEP'' \cite{materials_2016}); to test
\textbf{Correlation} in amorphous and liquid materials we selected
amorphous palladium (Rodríguez et al. \cite{rodriguez_emergence_2019}), amorphous palladium hydride
(Rodríguez \cite{rodriguez_calculo_2019}) and liquid bismuth. Because of the complexity to
calculate PAD of amorphous and liquids in Forcite, we chose to compare
them with the code developed by U. Santiago within our group. (Santiago
2011).

The results of these benchmarks are shown in \ref{fig:RDF-PAD}, and
the structures used to calculate these figures are included in the code
as tests 1 to 5. The last structure included as test 6 is a 2x2x2
supercell of amorphous palladium hydride included in test 4, to
benchmark memory and CPU performance in a structure with thousands of
atoms.

\section{Conclusion \& Perspective}

\textbf{Correlation} is a lightweight, modular software that can be used
in HPC and adapted to analyze the main correlation functions used to
characterize: crystalline, amorphous solids and liquids.

\textbf{Correlation} software has been used in previously published work
(Galván-Colín et al. \cite{galvan-colin_short-range_2015}; Galvan-Colin et al. \cite{galvan-colin_ab_2016}; Mata-Pinzón et al. \cite{mata-pinzon_superconductivity_2016}) as well as several PhD Theses (Santiago \cite{santiago_simulacion_2011}; Romero-Rangel \cite{romero-rangel_simulaciones_2014};
Mejía-Mendoza \cite{mejia-mendoza_estudio_2014}; Galvan-Colin \cite{galvan-colin_atomic_2016}; Mata-Pinzón \cite{mata-pinzon_propiedades_2016}; Rodríguez \cite{rodriguez_calculo_2019})
developed in our group. We will continue to support and enrich the
software in the foreseeable future. We are open to receive suggestions
that would further improve the functionality of the software. Address
all comments and observations to the first author, I.R:
isurwars@ciencias.unam.mx

\section{Acknowledgements}

I.R. acknowledges PAPIIT, DGAPA-UNAM for his posdoctoral fellowship.
D.H.R. acknowledges Consejo Nacional de Ciencia y Tecnología (CONACyT)
for supporting his graduate studies. A.A.V., R.M.V., and A.V. thank
DGAPA-UNAM for continued financial support to carry out research
projects under grants No.~IN104617 and IN116520. M. T. Vázquez and O.
Jiménez provided the information requested. A. López and A. Pompa helped
with the maintenance and support of the supercomputer in IIM-UNAM.
Simulations were partially carried out at the Supercomputing Center of
DGTIC-UNAM. We would like to express our gratitude to F. B. Quiroga, M.
A. Carrillo, R. S. Vilchis, S. Villarreal and A. de León, for the time
invested in testing the code, as well as the structures provided for
benchmarks and tests.

\section*{References}

\bibliographystyle{unsrt} 

\bibliography{paper.bib}

\end{document}